\documentstyle[aps,pra,epsfig,twocolumn]{revtex}

\def\be{\begin{equation}}
\def\ee{\end{equation}}
\def\bea{\begin{eqnarray}}
\def\eea{\end{eqnarray}}
\def\bma{\begin{mathletters}}
\def\ema{\end{mathletters}}
\def\C{\hbox{$\mit I$\kern-.7em$\mit C$}}
\def\c{{\cal C}}

\def\E{{\cal E}}

\begin{document}
\draft

\title{Entanglement molecules}

\author{W. D\"ur}

\address{Institut f\"ur Theoretische Physik, Universit\"at Innsbruck,
A-6020 Innsbruck, Austria}

\date{\today}

\maketitle

\begin{abstract} 
We investigate the entanglement properties of multiparticle systems, 
concentrating on the case where the entanglement is robust against disposal of 
particles. Two qubits ---belonging to a multipartite system--- are entangled in 
this sense iff their reduced density matrix is entangled. We introduce a family 
of multiqubit states, for which one can choose for any pair of qubits 
independently whether they should be entangled or not as well as the relative 
strength of the entanglement, thus providing the possibility to construct all 
kinds of ''Entanglement molecules''. For some particular configurations, we also 
give the maximal amount of  entanglement achievable.
\end{abstract}


\narrowtext
Entanglement is at the heart of Quantum Information theory. In recent years, 
there has been an ongoing effort to characterice quantitatively and qualitatively 
entanglenglement. While for bipartite systems this problem is essentially 
solved, it remains still open for multipartite systems. In this case, there 
exist several possible approaches to identify different kinds of multipartite 
entanglement, and many interesting phenomena related to multipartite 
entanglement have been discovered \cite{effects}. 

Here we concentrate on bipartite aspects of multipartite entanglement, in 
particular on bipartite entanglement which is robust against disposal of 
particles. We consider $N$ spatially separated parties $A_1,\ldots,A_N$, each 
possessing a qubit. We first investigate the $N$--party 
Greenberger-Horne-Zeilinger (GHZ)-state \cite{Gr89}
\be
|GHZ\rangle=\frac{1}{\sqrt{2}}(|0^{\otimes N}\rangle+|1^{\otimes N}\rangle) \label{GHZ},
\ee
which is considered to be a maximally entangled state (MES) of $N$ particles in 
several senses. For example, one can create a MES shared by any two of the 
parties with help of a (local) measurement performed by the remaining ones. Thus 
any two particles are potentially entangled, i.e. when allowing for assistance 
of the other parties, bipartite entanglement can be obtained from state 
$|GHZ\rangle$. However, it is essential that the remaining $(N-2)$ parties perform 
measurements to assist the other two parties to share entanglement. If however 
---for some reason--- the information about only one of the particles, say 
$A_N$, is lost (or party $A_N$ decides not to cooperate with the remaining 
ones), the state of the remaining parties is only classically correlated and thus not entangled. 
In particular, the reduced density operator of any two parties is 
separable\footnote{Given a $N$--partite state $\rho$, the reduced density 
operator $\rho_{12}$ of party $A_1$ and $A_2$ is defined as $\rho_{12}\equiv 
{\rm tr}_{3,\ldots,N} (\rho)$. The operator $\rho_{12}$ is separable if it can 
be written as a convex combination of product states.}. When considering the 
reduced density operator of two parties, we deal with the situation where the 
information about all remaining particles is not accessable (or the remaining 
parties are not willing to cooperate). 

We thus say that two particles are entangled iff their reduced density operator 
is non--separable, i.e. the two particles share entanglement, independent what 
happens to the remaining particles. Such a definition is very suitable from a 
practical point of view, as there are certain multipartite 
scenarious where one is interested in entanglement properties of pairs of 
parties, which are independent of other parties. Note that in this 
sense, the state $|GHZ\rangle$ contains no (bipartite) entanglement at all.

But are there $N$--particle states which are still entangled when tracing out 
{\it any} $(N-2)$ particles, i.e. are there states where all particles are 
entangled with all other particles ? And if this is possible, what is the 
maximal amount of entanglement the remaining two parties can share ? In this 
paper, we will answer these questions and we will consider the more general 
setup where some parties are entangled, while some others are not. For example, 
for $N=3$ one may have that the reduced density operators $\rho_{12}$ and 
$\rho_{13}$ remain entangled, but $\rho_{23}$ is separable. We will show that 
one can have all possible configurations of this kind, i.e. there exist states 
where one can choose for each of the reduced density operators $\rho_{kl}$, $k<l 
\in\{1,\ldots, N\}$ independently whether it should be entangled or not. This 
allows to build general structures of $N$ particle states, which we call 
'Entanglement molecules' in spirit of the generalization of Wootters idea of an 
'Entangled chain' \cite{Wo00}, where one has a string of qubits, each qubit 
being entangled only with its nearest neighbours.  We are considering more 
general setups, e.g. closed rings of particles where one only has nearest 
neighbour entanglement, (finite) strings with distance dependent entanglement 
(also 2nd and 3rd and so on neighbourhood entanglement), entanglement buckiballs 
(like the $C_{60}$ molecule), or more generally all possible setups of this kind 
one may imagine.

\begin{figure}[ht]
\begin{picture}(230,280)
\put(5,65){\epsfxsize=230pt\epsffile[26 637 223 832]{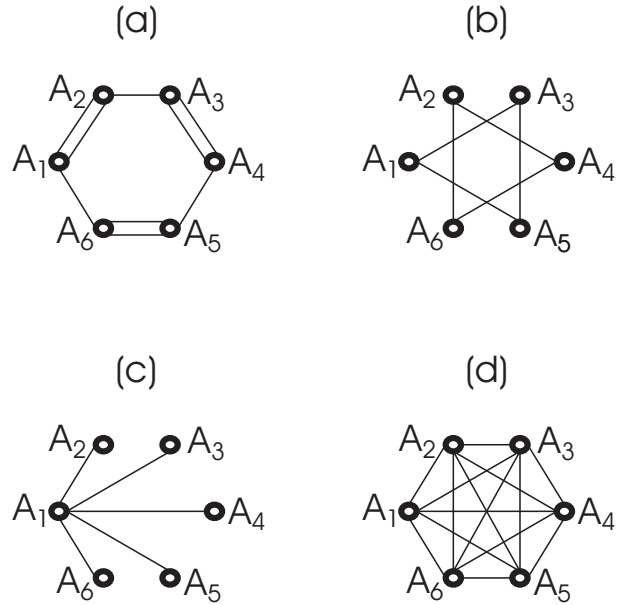}}
\end{picture}
\caption[]{Several kinds of Entanglement molecules for $N=6$. (a) Entanglement ring - double lines indicate stronger entanglement between particles. (b) All even [odd] particles are entangled respectively. (c) One particle $A_1$ equally entangled with five others. (d) All particles equally entangled.}
\label{Fig1}
\end{figure}

There are $N(N-1)/2$ different bipartite reduced density 
operators $\rho_{kl}$ which may be either separable or not. If the reduced density operator 
$\rho_{kl}$ is non--separable, this automatically implies that a MES shared 
between parties $A_k$ and $A_l$ can be distilled (when allowing for several 
copies of the state), even without help of the remaining parties. This is due to 
the fact that for two qubit systems, inseparability is equivalent to 
distillability \cite{Ho97}. In fact, the remaining parties can by no means 
prevent parties $A_k$ and $A_l$ to distill a MES.  In a diagram, this will be 
visualized by a line between particles $A_k$ and $A_l$ which represents 
entanglement between the two parties in question (see Fig. \ref{Fig1}). Each of 
the particles may be entangled with one (or more) of the remaining $(N-1)$ 
particles. In particular one can have that any particle is entangled with all the 
remaining ones. Clearly, it is interesting to ask how strong these 'bindings' 
(the entanglement between two particles) can be. Therefore, one has to quantify 
the entanglement of the bipartite reduced density operators $\rho_{kl}$. In this 
work, we choose as a measure of entanglement the concurrence $\c$ (for a 
definition of the concurrence see e.g. \cite{Wo99}). On one hand, we follow the 
lines suggested in \cite{Wo00,Wo99}, on the other we have that the entanglement 
of formation ---the amount of entanglement required to prepare a state $\rho$---  
is monotonically increasing with the concurrence $\c$ \footnote{The entanglement of 
formation is given by $E_f(\rho)=h(\frac{1}{2}+\frac{1}{2}\sqrt{1-{\cal C}^2})$, 
where ${\cal C}$ is the concurrence and $h$ is the binary entropy function 
$h(x)=-x{\rm log}_2x-(1-x){\rm log}_2(1-x)$.} and thus the concurrence itself 
may be used to measure the strength of the bindings. For two special cases of 
particular interest, we will give the states with the maximal achieveable 
strength of the bindings.



Let us start by introducing a family of $N$ qubit states, which includes all possible 
configurations of 'Entanglement molecules'. First we specify for each of the 
reduced density operators $\rho_{kl}$ whether it should be distillable or not, 
i.e. whether entanglement between the parties $A_k$ and $A_l$ can be distilled 
--- without help of the remaining parties --- or not. Let $I=\{k_1l_1, \ldots 
,k_Ml_M\}$ be the set of all those pairs where distillation should be possible, 
i.e. for $kl\in I$, we have that $\rho_{kl}$ is distillable. We define the 
state 
\be 
|\Psi_{ij}\rangle \equiv |\Psi^+\rangle_{ij} \otimes |0\ldots 0\rangle_{\rm rest}, 
\ee 
that is the particles $A_i$ and $A_j$ are in a MES, namely 
$|\Psi^+\rangle=1/\sqrt{2}(|01\rangle+|10\rangle)$, and the remaining particles 
are disentangled from each other and from $A_iA_j$. We now introduce a family of 
states which has the desired properties: 
\be
\rho_I= \frac{1}{M} \sum_{kl \in I}  x_{kl}|\Psi_{kl}\rangle\langle\Psi_{kl}|,\label{family}
\ee
where $M \equiv \sum_{kl \in I} x_{kl}$ is a normalization factor. It is 
straightforward to calculate the reduced density operators 
$\rho_{kl}$. In the standard basis, one can check for $kl \in I$ that $\rho_{kl}$ is of the form
\be
\rho_{kl}= \frac{1}{M}\left( \begin{array}{cccc} a & 0 & 0 & 0 \\ 
0 & b & x_{kl}/2 & 0 \\
0 & x_{kl}/2 & c & 0 \\
0 & 0 & 0 & 0 \end{array}\right), 
\ee
while for $mn \notin I$ one obtains
\be
\rho_{mn}= \frac{1}{M}\left( \begin{array}{cccc} \tilde a & 0 & 0 & 0 \\ 
0 & \tilde b & 0 & 0 \\
0 & 0 & \tilde c & 0 \\
0 & 0 & 0 & 0 \end{array}\right), 
\ee
It is simple to calculate the concurrence in both cases:
\bea
{\cal C}_{kl}&=& \frac{x_{kl}}{M} \mbox{ iff } kl\in I \nonumber\\
{\cal C}_{mn}&=& 0 \mbox{ iff } mn\notin I,
\eea
where we denote $\c_{kl}\equiv\c(\rho_{kl})$. We have that in $\C^2\otimes \C^2$ 
systems, nonzero concurrence automatically implies distillability of the state, 
while zero concurrence implies separability. Thus it is already clear that 
$\rho_I$ has the desired properties, i.e. entanglement between $A_k$ and $A_l$ 
can be distilled iff $kl \in I$. We see that we can arbitrarily choose the 
relative strength of the bindings (measured by the concurrence) via the positive 
coefficients $x_{kl}$. It is now straightforward to explicitly construct the 
examples illustrated in Fig. \ref{Fig1}: In (a) we have that 
$x_{12}=x_{34}=x_{56}=2/9, x_{23}=x_{45}=x_{16}=1/9$ (entanglement ring); In (b) 
$x_{kl}=1/6$ iff both $k$ and $l$ are even [odd] respectively and zero otherwise 
(all even [odd] particles are equally entangled)  ; In (c) we have that 
$x_{1l}=1/5$, while all other $x_{kl}=0$ (one particle equally entangled with 
five other ones); Finally, in (d) $x_{kl}=1/15$ (all particles equally 
entangled). In a similar way, one can construct all examples mentioned in the 
introduction, such as strings with distance dependent entanglement or 
entanglement buckiballs.

Note that in this construction, we have that 
\be
\sum_{i<k}\c_{ik}=1,
\ee
which shows that in a situation where all bindings have the same strength, the 
concurrence of the corresponding reduced density operators is determined by the 
total number of bindings, i.e. the number of elements in $I$. Hence, a state 
constructed in this way, where each particle is entangled with all others (i.e. 
all possible reduced density operators are distillable) and the strength of all 
bindings is equal has ${\cal C}_{kl}=2/[N(N-1)]$. As we shall see next, this is 
however not the maximum value one can achieve for this particular configuration.  
In the following, we will analyze the two situations illustrated in Fig. 
\ref{Fig1}(c) and (d), namely ---(i) all particles are equally entangled and (ii) 
one particle which is equally entangled with $(N-1)$ others--- and determine the 
maximum strength of the bindings for $N=3$.  

{\bf (i) All particles pairwise entangled}
\\
We start with the case where all particles are equally entangled with 
each other (see also Fig. \ref{Fig1}(d)). As shown in \cite{Du00}, the state
\be
|W\rangle=1/\sqrt{3}(|001\rangle+|010\rangle+|100\rangle) \label{W},
\ee
is the state of three qubits whose entanglement has the highest degree of 
endurance against loss of one of the three qubits. In particular, $|W\rangle$ 
maximizes the function
\be
\E_{\min}(\psi) \equiv \min(\c_{12}, \c_{13}, \c_{23}),
\label{worstcase}
\ee 
and has $\c_W\equiv\c_{12}=\c_{13}=\c_{23}=2/3$. It follows that 
$|W\rangle$ is the 3--qubit state where all particles are equally entangled and 
the entanglement --- measured by the concurrence $\c$--- is maximal. Comparing 
$|W\rangle$ with $\rho_I$ of the form (\ref{family}) with $I=\{12,13,23\}$, one 
immeadetly observes that $\c_{\rho_I}=1/3$, while $\c_{W}=2/3$.

More generally, let us consider the $N$--party form $|W_N\rangle$ of the state 
$|W\rangle$, defined as 
\be
|W_N\rangle \equiv 1/\sqrt{N}|N-1,1\rangle,
\ee
where $|N-1,1\rangle$ denotes the (unnormalized) totally symmetric state 
including $N-1$ zeros and $1$ ones, e.g. 
$|2,1\rangle=|001\rangle+|010\rangle+|100\rangle$. As shown in \cite{Du00}, state 
$|W_N\rangle$ is a $N$ qubit state with all reduced density operators equal and 
the concurrences are given by $\c_{kl}=2/N$, which has to be compared to ${\cal 
C}_{\rho_I}=2/[N(N-1)]$, $\rho_I$ being a state of the family (\ref{family}) 
where all particles are equally entangled. It is however not known whether 
$\c_{kl}=2/N, ~\forall k\not=l$ is the maximal value achievable.

{\bf(ii) One particle entangled with $N-1$ others}
\\
We consider now the case where one particle is equally entangled with $(N-1)$ others 
and determine the maximal possible strength ${\cal C}$ of the bindings (see Fig. 
\ref{Fig1}(c)). A related problem of this kind, namely the question of optimal 
entanglement splitting, i.e. the optimal way for a party $B$ to equally 
distribute its initial entanglement (shared with a party $A$) among several 
partners, was recently analyzed by Bru{\ss} in \cite{Br99}.

As shown in \cite{Wo99}, we have for $N=3$ that ${\cal C}_{12}^2 + {\cal 
C}_{13}^2  \leq 1$, i.e for ${\cal C}_{12} = {\cal C}_{13} \equiv {\cal C}$ we 
have that ${\cal C} \leq 1/\sqrt{2}$. This value is achieved by the state 
\be
|\psi\rangle=\frac{1}{\sqrt{2}}|100\rangle + \frac{1}{2}(|001\rangle + |010\rangle)\label{maxi12}.
\ee
More generally, it was conjectured in \cite{Wo99} that the inequality 
$\sum_{k=2}^N {\cal C}_{1k}^2 \leq 1$ also 
holds. If we demand again that ${\cal C}_{1k} = {\cal C}_{1l} \equiv {\cal C} 
\mbox{ } \forall k,l>1$, we find that ${\cal C} \leq 1/\sqrt{N-1}$. This value 
is obtained by the state
\be
|\psi\rangle=a|1\rangle|00\ldots 0\rangle + b |0\rangle|N-2,1\rangle)\label{form},
\ee
with $a=1/\sqrt{2}$ and $b=1/\sqrt{2(N-1)}$, where $|N-2,1\rangle$ is again an (unnormalized) 
totally symmetric state including $(N-2)$ zeros and one 1. For a state of the form 
(\ref{family}), where one particle is equally entangled with $(N-1)$ others, one 
obtains $\c_{1k}=1/(N-1)$.

For practical purposes, it may sometimes be useful to consider the fidelity of 
the reduced density operator (i.e. the maximal overlap of the reduced density operator
with any MES) ---which is generally not a proper measure of entanglement--- 
instead of the concurrence. The fidelity of the reduced density operator however 
indicates the achievable quality to perform certain quantum information tasks, 
e.g. teleportation \cite{Be93}. The fidelity of a density operator $\rho$ is 
defined as
\be
F_{\rho}=\mbox{max} \langle\Phi|\rho|\Phi\rangle,
\ee
where the maximum is taken over all maximally entangled states $|\Phi\rangle$. 
In the situation we consider, where one particle is equally entangled (i.e. 
$F_{\rho}$ is equal for all reduced density operators $\rho_{1k}$) with $(N-1)$ 
others, one can derive a bound for the maximum fidelity $F_{\rho}$ of the 
reduced density operators $\rho_{1k}$ using results from optimal cloning 
\cite{We98}. Given a density operator $\rho$ with a certain fidelity $F_{\rho}$, 
one can use $\rho$ to teleport \cite{Be93} the (unknown) state of a particle. As 
shown by the Horodecki \cite{Ho98}, one can find a teleportation protocol which 
works equally well for all input states and has the maximum teleport fidelity 
$F_t=(2F_{\rho}+1)/3$. In our situation, we have that particle $A_1$ is 
entangled with $(N-1)$ other particles. Thus, when performing a certain 
teleportation protocol \cite{Mu99,Du99JMO}, one obtains $(N-1)$ (imperfect) 
clones of a state at the locations $A_2,\ldots,A_N$, where the quality of the 
clones is determined by the teleport fidelity $F_t$ of the corresponding reduced 
density operators. As shown by Werner \cite{We98}, the maximum cloning fidelity 
of a $1 \rightarrow (N-1)$ cloner is given by $F_c=[2(N-1)+1]/[3(N-1)]$, from 
which follows that the teleport fidelity $F_t$ of the reduced density operators 
must fulfill $F_t \leq F_c$, otherwise one could construct in this way a cloning 
machine which works better than the optimal one, which is clearly impossible. We 
thus have that 
\be 
F_{\rho} \leq \frac{1}{2}+\frac{1}{2(N-1)}.
\ee
For the state $|\psi\rangle$ which maximizes the concurrence, we find that it 
does not obtain the maximal possible fidelity $F_\rho$. However, there exist 
states for which the maximal possible value of the fidelity is obtained. For 
example, a state of the form (\ref{form}) with $a=1/\sqrt{N(N-1)}$ and 
$b=\sqrt{(N-1)/N}$ has the desired properties, as well as the 'telecloning 
state' introduced in \cite{Mu99}. On the other hand, a state of the form 
(\ref{family}), where one particle is equally entangled with $(N-1)$ others has 
$F_{\rho}=[2+N]/[4(N-1)]$, much smaller than the optimal value. 

Note that the maximum value of the 
entanglement (measured by the concurrence) of a given qubit with its neighbours 
is not simply determined by the number of entangled neighbours, but also by the 
properties of the neighbours (i.e. the number of particles to which the 
neighbouring particles are entangled). This can be seen by noting that (i) in 
the case where one particle is equally entangled with two others (which are 
disentangled among themselves), the maximimal value for the concurrence is given 
by $\c=1/\sqrt{2}$ and (ii) in the case where three qubits are equally 
entangled, the maximum value is given by $\c=2/3$, which shows that the 
entanglement between systems $A_2$ and $A_3$ influence the maximum value of the 
entanglement between systems $A_1A_2$ and $A_1A_3$.

In summary, we provided a family of states which allows to construct all 
possible kinds of 'Entanglement molecules', i.e. one can choose for any pair of 
qubits independently whether a MES can be distilled without help of the remaining 
parties or not. In addition, the relative strength of the bindings (measured by 
the concurrence) can be adjusted arbitrarily. We investigated two particular 
configurations closer and provided states achieving the maximum value for the 
strength of the bindings.

I would like to thank G. Vidal and J. I. Cirac for discussions. This work was 
supported by the Austrian Science Foundation under the SFB ``control and 
measurement of coherent quantum systems´´ (Project 11), the European Community 
under the TMR network ERB--FMRX--CT96--0087 and project EQUIP (contract  IST-1999-11053), the European Science Foundation, 
and the Institute for Quantum Information GmbH.



\end{document}